\documentclass[preprint,12pt]{aastex}
\usepackage{epsfig}
\usepackage{apjfonts}

\begin{document}
\title{A Search for Asteroids, Moons, and Rings Orbiting White Dwarfs} 
\author{Rosanne Di\thinspace Stefano$^1$, 
Steve B. Howell$^2$, 
Steven D. Kawaler$^3$}  
\affil{$^1$ Harvard-Smithsonian Center for Astrophysics, 
60 Garden Street, Cambridge, MA 02138}
\affil{$^2$ National Optical Astronomy Observatory, 
950 N. Cherry Ave, Tucson, AZ 85719}   
\affil{$^3$ Department of Physics and Astronomy, Iowa State 
University, Ames, IA 50011}


\begin{abstract} 
Do white dwarfs host asteroid systems?
Although several lines of argument
suggest that white dwarfs may be orbited by large
populations of asteroids, 
transits would provide the
most direct evidence.
We demonstrate that the {\it Kepler} mission
has the
capability to detect transits of white dwarfs by asteroids.  
Because white-dwarf asteroid systems, if they exist,
are likely to 
contain many asteroids orbiting in a
spatially extended distribution, 
discoveries of asteroid transits can be made by  
monitoring only a small number of white dwarfs, compatible with 
{\it Kepler's} primary mission, which is to monitor stars
with potentially habitable planets.
Possible future missions that survey ten times as many stars with 
similar sensitivity and minute-cadence monitoring
can establish
the characteristics of asteroid systems around white dwarfs,
such as the distribution of asteroid sizes and semimajor axes.
Transits by planets would be more dramatic, but the probability 
that they will occur is lower. 
Ensembles of planetary moons and/or the presence of
rings around planets can also
produce transits detectable by {\it Kepler}.  The
presence of moons and rings can
significantly increase the
probability that {\it Kepler} will discover planets orbiting white
dwarfs, even while monitoring only a small number of them.   
\end{abstract}

\section{Introduction} 

Stars are orbited by dust, asteroids, planets and their moons.
As each star evolves, its planetary system evolves as well, 
through a combination of stellar expansion, mass loss, and
dynamical interactions. In spite of the fact that the first planet to be
discovered orbits a pulsar (Wolszczan \& Frail 1992), 
we still know little about 
the planetary systems around 
stellar remnants. Fortunately, a variety of methods are poised to
change this. Pulsar timing studies in combination with HST images have 
found a planet in a circumbinary orbit around a binary consisting of
a millisecond pulsar and a white dwarf (Sigurdsson et al.\, 2003).
Timing measurements of pulsating compact stars can also identify candidate 
planetary systems around highly evolved stars (i.e. Silvotti et al. 2007), 
with complementary {\it Spitzer} observations 
able to detect or place limits
on possible planetary companions of white dwarfs (Mullally et al.\, 2009).
In this paper we point out that the {\it Kepler} mission 
has the sensitivity and cadence needed to detect
transits of white dwarfs 
by asteroids. 

\subsection{Motivation}

Calculations show that asteroid or cometary systems can survive stellar 
evolution (Alcock et al.\, 1986).
While asymmetries in the mass loss can influence 
survivability, some white dwarfs could experience
asteroid impacts at a rate of $10^{-4}$~yr$^{-1}$ (Parriott and Alcock 1998).   
Several independent lines of evidence suggest that some white dwarf stars
do host circumstellar material ranging from dust to asteroids and
planets (Howell et al.\, 2008; Jura 2008 and references therein;
Farihi et al.\, 2009; Jura, et al.\, 2009; Jura, Farihi, \&
Zuckerman 2009).  
Asteroid-sized objects, with
diameters ranging from several tens of kilometers
up to and including dwarf planets, are the primary focus 
of this paper.
Their instantaneous orbital distance from the white dwarf ranges from
the tidal limit out to the equivalent of the Sun's Oort Cloud.
We will also consider the possible effects of rings and moons
orbiting planets in white-dwarf systems. 

White dwarfs with infrared excesses have been studied by a
number of groups. A common conclusion is that circumstellar
dust is present.
These white dwarfs tend to exhibit unusually strong metal lines
in their
photospheres (von Hippel et al. 2007),
perhaps showing the signs of enrichment by recent
accretion of material from an asteroid.
Indirect evidence of planetary material around warmer white dwarfs
includes several stars that show metal lines in their photospheric
spectrum (Zuckerman et al. 2007, and references therein).  In
white dwarfs,  metals sink below the photospheres on extremely short
time scales (of order days), so the presence of elements such as
calcium argues that the stars have recently accreted metal-rich
material.

As an example of possible asteroidal material surrounding a white
dwarf, consider the recent work by Gansicke et al. (2007).  They
report observations of
CaII and FeII double-peaked emission lines, interpreted to be from a
circumstellar, metal-rich gaseous disk.  The white dwarf itself is hot
enough ($T_{\rm eff} \sim 22000 K$) to burn off dust from the disk, accounting for
the
lack of an infrared excess.  The disk itself is the likely remnant of a
tidally disrupted rocky body of asteroid-sized mass (Gansicke et al.
2007).
The case for a metal-rich disk is strengthened by observations of MgII
absorption lines in the stellar spectrum.
Dynamical modeling of the system by Gansicke et al. (2007) constrain the
outer edge of the disk to be at about $1.2 R_{\odot},$ and places the
inner edge at approximately $0.64 R_{\odot}.$

Zuckerman et al. (2007) provide an analysis of the  white dwarf GD
362, including an estimate of the abundance of 17 elements accreted by
that star.  They conclude that an asteroid-mass object (either a
remnant asteroid or the residual of a disrupted terrestrial planet) is
needed to explain the abundance pattern.  A very rough estimate of the
size of an object needed to account for the metals seen in that star
($10^{22}$~g) is of order $80$~km; such an object would produce a transit
with a depth of $0.01\%$, or $100$~ppm.

\subsection{Goals}
For every asteroid that is tidally disrupted, there must be many more
with perihelia located much farther from the white dwarf.
Direct detection of these asteroids is challenging.
They have very little gravitational influence on their star
and cannot presently be detected through either
Doppler or ground-based photometric transit methods. 
Transits of white dwarfs by 
large asteroids can, however, be detected by {\it Kepler},
a space mission designed to detect the transits of Sun-like stars
by Earth-like planets. 
Consider a
white dwarf with a radius of $8000$~km. An asteroid with a radius of
$100$~km ($1000$)~km will produce a fractional decrease in the amount
of light received of 150~ppm (15,000~ppm), within Kepler's
detection limit. Many such asteroids of this size are known in our
own solar system [an
estimated $80,000$ in the Kuiper Belt alone (Trujillo, Jewett, \& Luu 2001)],
so it is reasonable to expect that
they exist elsewhere as well.

In \S 2  we show that the {\it Kepler} observatory can
discover asteroids orbiting white dwarfs by
identifying short-lived downward deviations from the baseline
flux in white dwarfs associated with transits by asteroids.
In \S~3 we discuss what we can learn through {\it Kepler} monitoring
of a small set
of white dwarfs.
Asteroid transits or significant limits on white-dwarf asteroid
systems are a certain science return. In addition,
depending on the structure of white-dwarf planetary systems, 
transits by rings and/or moons may also be detected
by monitoring a modest number of white dwarfs.  

\section{Kepler Detections of Asteroid Transits} 

\subsection{Detection of Transits}

The depth of a transit and its time duration determine its level of
detectability. 
If $A_{\rm ast}$ is the projected area of the asteroid, and
$A_{\rm wd}= \pi\, R_{\rm wd}^2$ the
cross-sectional area of the
white dwarf, the depth of the transit is $(A_{\rm ast}/A_{\rm wd}).$
The calculations for asteroids transiting
white dwarfs mirror the results for an Earth-like planet
transiting a Sun-like star, because the relative size scales are similar.
For an asteroid of a given size, the depth is greatest for
more massive white dwarfs, which are smaller.

Asteroid transits against a white dwarf will be of short
duration and may have distinctive profiles.
The time required for the asteroid's center of mass to cross the
diameter of the white dwarf is $\tau_{\rm cross} = 2\, R_{\rm wd}/v,$
where $v$ is approximately equal to the orbital velocity.
\footnote{Equation~1 neglects the size of the asteroid, the proper motion of the
white dwarf, and assumes that the transit occurs along a diameter
of the disk.}.
\begin{equation} 
\tau_{\rm cross}=9.4\, {\rm minutes}\, 
            \Big({{R_{\rm wd}}\over{7.5 \times 10^8 {\rm cm}}}\Big)\, 
\Big({{a}\over{{\rm AU}}}\Big)^{{1}\over{2}}  
\Big({{0.8\, M_\odot}\over{M_{\rm wd}}}\Big)^{{1}\over{2}}  
\end{equation} 
Because the orbital speed, $v,$ decreases with increasing $a$,
an asteroid of fixed size produces a longer event when it is farther
from the white dwarf.
The crossing time is larger for less massive white dwarfs because
the orbital speed is smaller (for a given $a$)
and because the radius of a low-mass white
dwarf is larger. 
The ingress and egress may be distinctive because many asteroids will
not be massive enough to have been pulled into a spherical shape
by self gravity.  

We have carried out a set of calculations to determine how large an
asteroid must be in order for its transit against a white dwarf
of given brightness to be detectable by {\it Kepler.}
We use the information on the {\it Kepler}
web pages\footnote{http://kepler.nasa.gov/}
to estimate the number of detected photoelectrons per minute:
$N = 1.3\times 10^7 \, (10^{-0.4 (M_{Kepler}-12)})$~per~minute.
For representative examples,
we used the {\it Kepler} magnitudes of $3$ relatively bright white dwarf
candidates in
the {\it Kepler} field:
$M_{Kepler} = 12.2, 14.8, 15.8.$
For each candidate white dwarf we carried out two sets of
calculations.  In the first set, we assumed that
the mass of the white dwarf was
$M_{\rm wd}=0.6\, M_\odot,$ and used the corresponding
radius. In the second
set we used $M_{\rm wd}=1.3\, M_\odot,$ and decreased the value of
$R_{\rm wd}$ accordingly.
Then, for each of $60$ values of the orbital separation $a$, we
computed the diameter of an asteroid for which we would have a $1 \sigma,$
$2 \sigma,$ and $3 \sigma$ detection of the transit, by integrating
over the crossing time. The results are shown in
Figure 1.
For lower-mass white dwarfs, asteroids of smaller
diameter can produce detectable transits.
For the brightest white dwarf, transits of asteroids in the
$100$-km class can be detected even when they are relatively close to
the white dwarf. The dimmer the white dwarf, the farther from it
must a $100$-km asteroid be in order for its transit to be
detected by {\it Kepler.} In the solar system, the Kuiper Belt
extends from roughly $30$~AU to $50$~AU. Large interlopers from the Oort Cloud,
such as 2006 SQ372,
are also found in this region, while the bulk of the Oort cloud lies beyond
$1000$~AU. Although the white dwarf systems we target for {\it Kepler}
study may
be very different, the example of the solar system indicates that it is good
to be sensitive to asteroids at large values of $a$.
\begin{figure*}[h]
\center{\includegraphics[width=7.5 cm,
height=12.5 cm,angle=-90]{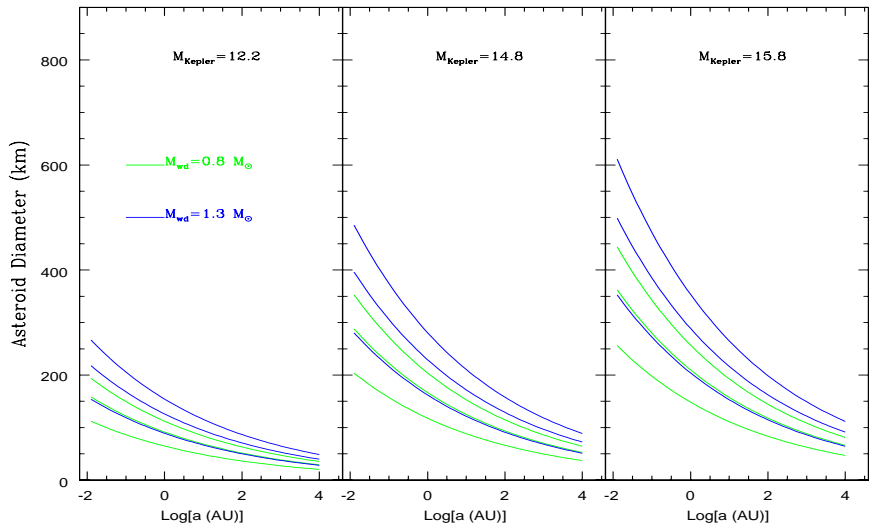}}
\caption{\footnotesize
{
The asteroid diameter (in km) necessary in order to detect a transit vs the
orbital separation (in AU) at the time of transit. The significance of the
detection was estimated
by integrating over the crossing time. The mass of the white dwarf was taken to
be $0.6\, M_\odot$ in the green (lighter) curves and $1.3\, M_\odot$ in the
blue (darker) curves. For each white dwarf, the (lowest, middle, top)
curve corresponds, respectively,
to a ($1 \sigma,$ $2 \sigma,$ $3 \sigma$) detection.
Each panel refers to the {\it Kepler} magnitude
of a specific candidate white dwarf known to
be in the {\it Kepler} field.
}}
\label{fig1}
\end{figure*}

\subsection{The Numbers of Asteroids}

The probability of detecting a transit by an
individual asteroid is small, because
it is proportional to the maximum angle of orbital inclination for which
a  transit can be observed: $(R_{\rm wd} + R_{\rm ast})/a,$ where
$R_{wd}$ is the radius of the white dwarf, $R_{ast}$
is the radius of the asteroid, and $a$ is the orbital separation.
If a typical value of $(R_{\rm wd} + R_{\rm ast})$ is $8 \times 10^8$~cm,
and the average value of $a$ at the time of transit
is $1$~AU, then taking inclination alone into account, we would have to
observe almost $30,000$ white dwarfs to have a good chance of
detection.\footnote{In this discussion of orientation
effects we neglect duty cycle issues, but note that if the orbital
period is large compared to the total time during which observations
occur, there is a further diminuation of the detection probability.}
If each white dwarf is orbited by many asteroids,
the probability of detection increases. Should all of the
asteroids orbit in a common plane, however,
we must still monitor a large number
of white dwarfs to have a good chance of detection.

Fortunately, our own solar system offers hope, and star and planet
formation theory also suggest that
planetary systems each host large numbers of asteroids and that
the orbits are not aligned. The Oort Cloud ($a > 1000$~AU), which
may have as much as $100$ Earth masses
(Marochnik et al.\, 1988)
is expected to have a
nearly spherical distribution. Some of its members have perihelia
within the Kuiper Belt ($30-50$~AU), while others could make even closer
approaches.
If each white dwarf we observe has a distributions of asteroids with the
geometry of the Oort Cloud, we have a good chance to detect transits
with {\it Kepler} monitoring of even a single white dwarf.
The Kuiper Belt itself has a scattered component
in which
the average orbital inclination angle, $i,$ is $12^o,$ while individual
orbits can be even more inclined. (See Sheppard 2006 and
references therein.) Therefore, if the asteroid systems
of white dwarfs have a geometry similar to that of the scattered Kuiper Belt,
and if we could monitor $10-15$ white dwarfs,
we would have a good chance that the equivalent of the scattered
``Kuiper'' Belt of one or more of them was inclined toward our line of sight.

Assuming a spherical distribution,
we can quantify the probability of detecting a transit as a function
of $a$ as follows. We compute the number of asteroids that would have to
have periastrons at a particular value of $a$ in order to have a
probability near unity of detecting the transit.
Given $(R_{\rm wd} + R_{\rm ast}) = 8 \times 10^8$~cm,
then for
$a=$~(0.1, 1, 10, 100)~AU,  the number of such asteroids is
($1.9\times 10^3$, $1.9\times 10^4$, $1.9\times 10^5$, $1.9\times 10^6$). 
These numbers assume that the interval $T$ during which continuous observations 
occur spans the orbital periods of the asteroids. In fact, $T$ will be
longer than $P_{orb}$ for small orbital apastrons, and the scaling above
holds for those separations. Even better, for $T>n\, P_{\rm orb}$, $n$ transits
will be detected; confidence that the photometric
dips were caused by transits can therefore be high, 
just as multiple planetary transits enhance
confidence in the discovery of planets.     
For wider separations, however, the probability is reduced by a factor
$T/P_{\rm orb}.$ For circular orbits, the number of asteroids needed to have   
a transit detection probability near unity is 
\begin{equation}  
N = 1.8 \times 10^4\, 
\Bigg(\frac{{\rm yr}}{T}\Bigg)\, \Bigg(\frac{M_\sun}{M}\Bigg) 
\Bigg(\frac{R_{\rm wd} + R_{\rm ast}}{8 \times 10^8 \rm{cm}}\Bigg)
\Bigg(\frac{a}{{\rm AU}}\Bigg)^{\frac{5}{2}}. 
\end{equation} 
If $T=P_{\rm orb}$ for $a=1\, AU,$ then the numbers of asteroids needed to
ensure detection for $a=$~(0.1, 1, 10, 100)~AU are
($1.9\times 10^3$, $1.9\times 10^4$, $6.0\times 10^7$, $1.9\times 10^{11}$).  

These numbers are modest enough to suggest that,
unless the white dwarf systems are depleted in $100$-km class
asteroids relative to the solar system, {\it Kepler}
can discover asteroid transits by monitoring
a handful of white dwarfs.
Furthermore, these numbers above are small enough that a null result
would represent a meaningful limit.

\subsection{Interpretation: Individual Events}

The characteristics of transit light curves 
are related to the properties of the asteroid.\footnote{ 
Note that the white dwarf may be intrinsically variable. If the
variability is periodic, it will not interfere with our ability
to detect a transit. 
Nevertheless, when analyzing the light curves, 
care must be taken to consider the
influence of any intrinsic variability because 
if 
the variability is complex,
detectability may require a deeper, longer-lasting transit.
To simplify the in the text, we discuss
the flux ${\cal F}$
as if it is constant when a transit is not occurring.
}
If the ingress and egress can be
resolved in time, there will be an interval $\delta t_{in}$
during which the flux declines by $\Delta {\cal F}$,
a second interval $\Delta T,$ during which the flux remains
at its steady minimum value ${\cal F}-\Delta {\cal F}$, and a third interval
$\delta t_{out}$ during which the transit ends and the flux
returns to the level, ${\cal F}$ it would have had without the transit.

It is important to note that, particularly for the bright
white dwarfs likely to be monitored by {\it Kepler,}
the white dwarf's radius and mass can be estimated to high accuracy
using good quality optical spectra and model fits to $T_{\rm eff}$
and $log(g)$. 
The depth of the transit (the maximum downward deviation, $\Delta {\cal F}$),
therefore measures the projected area of the asteroid.
The value of $\Delta T,$ combined with the estimated 
white dwarf's radius,
provides an estimate of the asteroid's
speed, hence its distance from the white dwarf.
If $\delta t_{in}$ and $\delta t_{out}$ cannot be measured,
then we can use the time resolution of the observations to
place upper limits on the linear dimensions of the asteroid.
If they can be measured, then we can
(1) determine the projected linear size of the asteroid at the time of ingress
and (2) also at the time of egress. If they are different, then the
asteroid
may have been spinning; we can (3) check for consistency to determine if a
realistic spin period is consistent with the observed change.
If they are the same, we can (4) determine if the shape of the light
curve is
consistent with a disk-like structure. Using the area (from the depth
of the transit) to estimate the possible mass, we can (5) check if we expect
the asteroid to be spheroidal. Finally, we can (6) check if the
linear dimensions as estimated during ingress and egress are
consistent with the projected area, as estimated from the depth
of transit.

Note that if an event fails consistency checks,  we can rule
it out as a transit candidate. Passing the checks does not
however confirm the
transit interpretation. When transit candidates are identified, 
an exhaustive analysis is required of any effects that could have produced
a false positive signal. In an individual case, if the asteroid orbit
happens to not be highly eccentric and if $a$ is not much larger than an
AU, we may see a repeat during the lifetime of {\it Kepler;} this will
confirm the transit model.
Overall, it is likely that an asteroid
population large enough to produce a transit by one asteroid
will produce transits by several independent asteroids,
and that a set of self-consistent results will increase confidence
in the transit interpretation.

\subsection{Interpretation: Populations of Asteroids}

The nature of the results that can be obtained
by {\it Kepler} depends on the characteristics of planetary systems.
Because the progenitor of a white dwarf was a giant, it seems
likely that the region within roughly an AU was cleared of planets
and asteroids. Yet, the evidence sited in the introduction of this
paper indicates that asteroids can and do approach close to white dwarfs.

If the monitored white dwarfs have close-in asteroids with small
semi-major axes,
we will discover ``repeats''in a sense: multiple transits with similar
characteristics. If the monitored white dwarfs have large asteroids,
the photometric dips during transits will be highly significant.
If the monitored white dwarfs have $\sim 10^{12}$ asteroids in a
Kuiper-like belt, several events caused by different asteroids
will be observed.
Even if individual events are detected with low confidence
(e.g., $1-\sigma$ photometric dips), we
may be able to derive significant results when several such are detected.
This is because
the probability of detecting multiple
dips due to random processes
(which we can assess through observations of other stars)
is expected to be low.
Thus, multiple transits caused by one or by several asteroids, or
deep transits, or long-lasting transits
would
provide information
about some characteristics of the asteroid system.

It is certainly
possible, however,  that for one or more white dwarfs, no highly significant
events are discovered. In this case, given the estimated efficiency,
which will be well known based on {\it Kepler's} observations of
hundreds of thousands of other stars, we can place
quantitative limits on the presence of close-in, and/or large, and/or
numerous asteroids around any given white dwarf.
We therefore expect monitoring of each white dwarf to produce significant
results of either a positive or negative nature.

\section{Prospects}  

\subsection{Asteroids}
The {\it Kepler} team is about to announce the first year's results.
The mission is scheduled to
take data for $3.5$~years, and could extend operations for a
total duration of $5$~years.  {\it Kepler} can transmit data on
approximately $170,000$ targets, most main-sequence stars which
are being monitored in hope of detecting transits by Earth-like planets.
A limited number of data slots are available to monitor other targets
suggested by the community. Two modes of monitoring are available:
$30$-minute cadence and $1$-minute cadence. Equation~1 shows that the
$1$-minute mode is needed to detect transits by close-in asteroids.
Although asteroids making close approaches to the white dwarf must be larger 
if their transits are to be detectable, it is important to 
be sensitive to close approaches for two reasons. First, the probability
that the orientation is favorable scales as $1/a.$ Second, if
the orbit is circular, the transits could be periodic, allowing for 
repeated transit observations.  One-minute cadence is also important if we are
to resolve the transit light curve for more distant approaches.

The need for $1$-minute cadence limits the number of white dwarfs
that can be monitored. In addition, only
a small number of bright white dwarfs in the {\it Kepler}
field are known. Fortunately, the large numbers of asteroids
expected per white dwarf will almost certainly make it possible to discover 
asteroids by monitoring almost any white dwarf that has them. 

Because planets are likely to form in a bottom-up approach, stellar
formation seems likely to always produce small masses that will 
be gravitationally bound to the star, regardless of whether large 
planets form. Even though stellar evolution is associated with
mass loss from the system, a large number of asteroids should remain bound.
In addition, the dynamical evolution of planetary orbits during 
stellar evolution seems likely to yield collisions and additional space debris
in the form of asteroids. This line of argument is consistent
with the data summarized in the introduction, which argues independently
that asteroids orbit white dwarfs.
Nevertheless, some white dwarfs may be less likely to host asteroid systems,
at least the ones associated with planet formation. Consider,
for example, a white dwarf that
emerged from a common envelope episode. This suggests that the
white dwarfs most suitable for the first monitoring program are
those with masses near or above $0.6\, M_\odot;$ in addition,
they should not have close stellar companions.\footnote{White dwarfs
in close binaries and those that have been involved in prior mass transfer
may also be interesting, but the science to be explored in those cases
is different. We therefore suggest that a limited program focus on isolated
white dwarfs.}

If  white-dwarf asteroid systems occupy a region similar
to the Solar System's scattered disk, 
then by monitoring a set of white dwarfs,
we can sample a random distribution of possible orientations.
If therefore, {\it Kepler} 
can monitor (for approximately one year each) roughly a dozen
bright white dwarfs, it should discover asteroids and begin to
quantify the 
fraction of white dwarfs with asteroids in the $100$-km class.

Future projects that can take this study further are
under consideration.  Since missions with {\it Kepler}'s sensitivity
can detect asteroids around white dwarfs, a more comprehensive
all-sky survey monitoring $\sim 2.5\times 10^6$~stars (such as that 
proposed for the Transiting Exoplanet Survey Satellite {\it TESS})
will be able to establish the statistics of asteroid systems
around white dwarfs: the frequency as a function of white dwarf
properties, and the distributions of asteroid sizes and orbital
separations. 

\subsection{Planets, Moons, and Rings}

White dwarfs may well be orbited by planets, but the probability
${\cal P}$ that 
the orientation of a planetary orbit is favorable for the detection
of a transit is small.
\begin{equation}
{\cal P} =  {{(R_{\rm wd}+R_{\rm pl})}\over{a}}
                =  2.0\times 10^{-4} 
                     \Big({{R_{\rm wd}+R_{\rm pl}}\over
                                {3 \times 10^9 {\rm cm}}}\Big)\,  
                     \Big({{{\rm AU}}\over{a}}\Big).
\end{equation}
This implies that
thousands of white dwarfs would have to be monitored in order to
discover transits by planets, which is not compatible with 
the primary goal of the {\it Kepler} mission.    
Nevertheless,  other signatures of planets are
more likely to be detected
by {\it Kepler}.

Although planetary 
rings are generally composed of bits of debris that are
individually too small to produce detectable transits, the combined
effect can be to absorb and scatter enough light that the transit
of the ring
will be detectable (e.g., Ohta, Taruya, \& Suto 2009 and 
references therein). In this case, the ingress and egress
profiles are likely to be symmetric and distinguishable from
the patterns produced by an isolated mass. 
Strategies for seacrhing for evidence of rings and moons
in transit light curves have been developed (Barnes \& Fortney 2004; 
Barnes 2004). HST observations
of the planetary transit of HD 189733 were able to dervie a 
convincing null result, ruling out the presence of rings or moons 
around around HD 189733b through a detailed light curve analysis
(Pont et al.\, 2007).
{\it Kepler} could do the same, or else discover moons and
rings around white dwarf planets, should they exist.  

To compute the probability of detecting a transit by a ring,
the term $R_{\rm pl}$ in Equation~2 must be replaced by 
$R_{\rm ring}\, sin(\theta)$ In this expression  
$R_{\rm ring}$ is the outer diameter of the ring system, and can
be significantly larger than $R_{\rm pl}$. Thus, even though the 
planet itself may out of our line of sight, the ring
could transit.  In the case of Saturn, for example, the
outer radius of its ring system ($4.8 \times 10^{10}$~cm for the E~ring)
 is almost ten times larger
than the radius of the planet. 
The angle $\theta$ in Equation~2 is the angle between the plane of
the ring and the orbital plane of the planet.
If $\theta=0,$
then transits by the ring will only be detected in some cases
in which the planet transits as well. In those cases
the effect of the ring will be striking, because it
will produce a diminuation in light from the white dwarf that lasts
significantly longer than the transit by the planet.
If, however, the plane of the ring is oriented at a non-zero  
angle $\theta$ relative to the
orbital plane, then the probability of a transit by the rings
can be greater than the probability of a transit by the planet.
($R_{\rm ring}\, sin(\theta)) > (R_{wd}+R_{pl})$    
If a Saturn-like planet were to orbit a white dwarf at $a=0.1$~AU,
then with $\theta=90,$ the value of ${\cal P}$ would be 
$0.04.$ If such systems are common,
then a ring transit could be discovered by monitoring one to two
dozen white dwarfs.  

Planets are also orbited by moons.  
Our own solar system contains more than $150$
moons, many 
large enough to produce transits of a white dwarf that  would be
detectable by {\it Kepler.}
The cases in which a moon is likely to produce a transit, even 
if the planet does not transit, are those in which the 
planet orbits the star many times during the course of the
monitoring observations and in which the moon also orbits the
planet many times during the same interval. 
Let  $a_m$ be the distance between the planet and its moon, and
let $\theta$ be the angle between the orbital planes of the moon and
planet. 
To compute ${\cal P}$,
the term $R_{\rm pl}$ in Equation~2 must be replaced by 
$a_m sin(\theta).$   
Consider a Saturn-like planet orbited by a moon at  
$10^{11}$~cm (an  $11$~day orbit).
If the distance of the Saturn-like
planet from an $0.8\, M_\odot$ white dwarf is equal to $0.2$~AU
(a $36.5$~day orbital period), and $\theta=90^0,$ then 
${\cal P}=0.04$ As with the case of rings, the probability
of that a detectable transit will occur during a year of monitoring 
one to two dozen white dwarfs is significant.

Altough we do not have the
{\it a priori}  knowledge needed to 
assess the likelihood of transits by rings or moons, 
the discovery that
exoplanets commonly have properties that were unexpected  
leads us to consider a range of possibilities for planets
orbiting white dwarfs. The considerations above show that,
{\it if}`   
white dwarfs tend to be orbited by close-in 
planets, and i{\it if}` these
planets have rings and/or moons, there is a chance that
{\it Kepler} will discover them by monitoring a modest
number of white dwarfs.  
An all--sky survey with the sensitivity of {\sl Kepler} would either discover such
systems or definitively rule them out. 

Consider the possibility that  
white dwarfs host both asteroid systems
and close-in planets with
rings and/or moons.
Dynamical stability arguments place constraints on the
number of close-in planets and on the linear dimensions of 
the system of moons orbiting each. Unless, therefore,
the asteroid systems are deficient in large asteroids relative to
what we might expect based on the solar system, transits by
asteroids should provide the dominant signal.

\subsection{Other white-dwarf science}
The continuous monitoring
of white dwarfs can lead to significant scientific results
in addition to those associated with asteroids.
If coherent oscillations are present in these stars, astroseismic
analysis would reveal these modes at amplitudes of 16 ppm ($3 \sigma$)
in just one month of data on a 15th magnitude star, and 4.6 ppm
($3 \sigma$) in one year. These are 10-30 times (or more) lower than any
ground-based photometry has achieved.
At these low levels new
pulsating classes could well be discovered.

\acknowledgements
We thank the anonymous referee for very helpful suggestions which have helped us clarify the discussion and analysis presented in this paper.

\references
\smallskip

\noindent
Alcock, C., Fristrom, C. C., \& Siegelman, R. 1986, ApJ, 302, 462 

\smallskip

\noindent
Barnes, J.~W., \& Fortney, J.~J.\ 2004, \apj, 616, 1193 

\smallskip

\noindent
Barnes, J.~W.\ 2004, 
Ph.D.~Thesis,  

\smallskip

\noindent
Farihi, J., Jura, M., 
\& Zuckerman, B.\ 2009, ApJ, 694, 805

\smallskip

\noindent
Gansicke, B., Marsh, T.R., Southworth, J., \& Rebassa-Mansergas, A
2007, Science 314, 1908

\smallskip

\noindent
Giclas, H.L, Burnham, R., \& Thomas, N.G. 1967,
Lowell Observatory Bulletin, no. 141, p. 49

\smallskip

\noindent
Howell S. B., et al., 2008, ApJ, 685, 418

\smallskip

\noindent 
Jura, M., Muno, M.~P., 
Farihi, J., \& Zuckerman, B.\ 2009, \apj, 699, 1473 

\smallskip

\noindent 
Jura, M., Farihi, J., 
\& Zuckerman, B.\ 2009, \aj, 137, 3191 

\smallskip

\noindent
Jura, M. 2008, AJ, 135, 1785

\smallskip

\noindent
Lepine, S. \& Shara, M.M. 2005, AJ,
129, 1483

\smallskip

\noindent
Marochnik, L.~S.,
Mukhin, L.~M., \& Sagdeev, R.~Z.\ 1988, Science, 242, 547

\smallskip

\noindent
Mullally, F., Reach, 
W.~T., De Gennaro, S., \& Burrows, A.\ 2009, \apj, 694, 327

\smallskip

\noindent
Ohta, Y., Taruya, A., \& Suto, Y. 2009, ApJ, 690, 1 
\smallskip

\noindent
Parriott, J., \& Alcock, C.\ 1998, \apj, 501, 357 

\smallskip

\noindent
Pont et 
al.(2007)]{2007A\&A...476.1347P} Pont, F., et al.\ 2007, \aap, 476, 1347 

\smallskip

\noindent
Silvotti, R., et al. 2007, Nature, 449, 189

\smallskip

\noindent
Sigurdsson, S., 
Richer, H.~B., Hansen, B.~M., Stairs, I.~H., 
\& Thorsett, S.~E.\ 2003, Science, 301, 193

\smallskip

\noindent
Trujillo, C.A., Jewett, D.C., \& Luu, J.X. 2001, AJ, 122, 457

\smallskip

\noindent
Sheppard, S.S. 2006, ASPCS, 352,3

\smallskip

\noindent
von Hippel, T., Kuchner, M.J., Kilic, M., Mullally, F., \& Reach, W.T.
2007, ApJ, 652, 544

\smallskip

\noindent 
Wolszczan, A., \& Frail, D.~A.\ 1992, \nat, 355, 145 

\smallskip

\noindent
Zuckerman, B., Koester, D., Melis, C., Hansen, B., \& Jura, M. 2007,
ApJ, 671, 872

\end{document}